\documentclass[pdflatex, sn-nature, useAMS, usenatbib]{sn-jnl}%for submission

\usepackage{graphicx}%
\usepackage{multirow}%
\usepackage{amsmath,amssymb,amsfonts}%
\usepackage{amsthm}%
\usepackage{mathrsfs}%
\usepackage[title]{appendix}%
\usepackage{xcolor}%
\usepackage{textcomp}%
\usepackage{manyfoot}%
\usepackage{booktabs}%
\usepackage{algorithm}%
\usepackage{algorithmicx}%
\usepackage{algpseudocode}%
\usepackage{listings}%

\usepackage{hyperref}%

\newcommand{\apjl}{Astrophys. J. Lett.}   % Astrophysical Journal, Letters
\newcommand{\apjs}{Astrophys. J. Suppl. Ser.}   % Astrophysical Journal, Supplement
\newcommand{\ao}{Appl. Opt.}   % Applied Optics
\newcommand{\aap}{Astron. Astrophys.}   % Astronomy and Astrophysics

\jyear{2023}%

\begin{document}


\newcommand{\vmax}{$v_\mathrm{max}$ }
\newcommand{\vpeak}{$v_\mathrm{peak}$ }
\newcommand{\sibelius}{ \sc{Sibelius} }
\newcommand{\Ms}{ \mathrm{M}_\odot }
\newcommand{\MM}{ \mathrm{M} }
\newcommand{\LL}{ \mathrm{L} }
\newcommand{\LCDM}{$\Lambda$CDM }
\newcommand{\cMpc}{ \mathrm{cMpc} }

\title[Distinct distributions of elliptical and disk galaxies as a $\Lambda$CDM prediction]{Distinct distributions of elliptical and disk galaxies across the Local Supercluster as a $\Lambda$CDM prediction}

\author*[1,2]{\fnm{Till} \sur{Sawala}}\email{till.sawala@helsinki.fi}
\author[2]{\fnm{Carlos} \sur{Frenk}}
\author[3, 4]{\fnm{Jens} \sur{Jasche}}
\author[1]{\fnm{Peter H.} \sur{Johansson}}
\author[4]{\fnm{Guilhem} \sur{Lavaux}}

\affil[1]{\orgdiv{Department of Physics}, \orgname{University of Helsinki}, \orgaddress{\street{Gustaf H\"allstr\"omin katu 2}, \city{Helsinki}, \postcode{FI-00014}, \country{Finland}}}

\affil[2]{\orgdiv{Institute for Computational Cosmology}, \orgname{Durham University}, \orgaddress{\street{South Road}, \city{Durham}, \postcode{DH13LE}, \country{United Kingdom}}}

\affil[3]{\orgdiv{The Oskar Klein Centre, Department of Physics}, \orgname{Stockholm University}, \orgaddress{\street{Albanova University Center}, \city{Stockholm}, \postcode{106 91}, \country{Sweden}}}

\affil[4]{\orgdiv{CNRS \& Institut d'Astrophysique de Paris}, \orgname{Sorbonne Universit\'e}, \orgaddress{\street{98 B Bvd Arago}, \city{Paris}, \postcode{75014}, \country{France}}}

\abstract{Galaxies of different types are not equally distributed in the Local Universe. In particular, the supergalactic plane is prominent among the brightest ellipticals, but inconspicuous among the brightest disk galaxies. This striking difference provides a unique test for our understanding of galaxy and structure formation. Here we use the {\sc Sibelius Dark} constrained simulation to confront the predictions of the standard Lambda Cold Dark Matter ($\Lambda$CDM) model and standard galaxy formation theory with these observations. We find that {\sc Sibelius Dark} reproduces the spatial distributions of disks and ellipticals and, in particular, the observed excess of massive ellipticals near the supergalactic equator. We show that this follows directly from the local large-scale structure and from the standard galaxy formation paradigm, wherein disk galaxies evolve mostly in isolation, while giant ellipticals congregate in the massive clusters that define the supergalactic plane. Rather than being anomalous as earlier works have suggested, the distributions of giant ellipticals and disks in the Local Universe and in relation to the supergalactic plane are key predictions of the $\Lambda$CDM model.}

\maketitle
The Local Supercluster is the largest structure in the Local Universe \cite{deVaucouleurs-1956}, defining the supergalactic plane and the supergalactic coordinate system \cite{deVaucouleurs-1976, deVaucouleurs-1991}. While the plane was originally defined by the relatively nearby clusters such as Virgo and Fornax, it is now understood to extend to at least $z=0.02$ as an excess of bright ellipticals and radio galaxies near the supergalactic equator \cite{Shaver-1989,  Tully-1992, Strauss-1993}. Strikingly, however, there is no corresponding excess in bright disks \cite{Shaver-1991, Peebles-2022b}. 

In the standard paradigm of hierarchical structure formation \cite{White-1991}, denser regions are characterised by a greater abundance of dark matter haloes, in particular more massive ones \cite{Benson-2000}. At the same time, standard galaxy formation theory predicts that disk galaxies evolve largely in isolation, growing primarily through in-situ star formation fuelled by the continuous accretion of gas \cite{Governato-2007, Guedes-2011}. By contrast, the higher ambient temperature in denser regions restricts the gas supply into haloes, which leads to gas depletion and the eventual quenching of star formation. Denser regions are also characterised by a higher merger rate, and a higher fraction of gas-poor mergers, both of which are understood to lead to the formation of elliptical galaxies \cite{Naab-2009, Johansson-2012}.

However, the metamorphosis of galaxies is believed to take hundreds of millions of years, precluding its direct observation in individual objects. The differing distributions of galaxies of different types and masses in different environments within the Local Universe thus provides an important opportunity to simultaneously test the ability of the cosmological framework to explain the observed large-scale structure, and our understanding of galaxy formation theory.

\subsection*{A Local Universe constrained simulation}
Direct comparisons of model predictions to observations in the Local Universe require a constrained simulation \cite{Mathis-2002, Gottloeber-2010, Libeskind-2020, Sawala-2022, McAlpine-2022, Stopyra-2024}. In this work, we use the {\sc Sibelius Dark} simulation \cite{McAlpine-2022}, hereafter {\sc Sibelius}, that is designed to reproduce the galaxy distribution out to a comoving distance of $\sim 200$  Mpc ($z \sim 0.04$). The initial conditions are constrained using the Bayesian Origin Reconstruction from Galaxies ({\sc BORG}) algorithm \cite{Jasche-2013, Lavaux-2016, Jasche-2019} to represent the most likely initial conditions that give rise to the observed large-scale structure in the 2M++ galaxy redshift survey \cite{Lavaux-2011}, in the $\Lambda$CDM cosmology, with parameters $\Omega_\Lambda = 0.693$, $\Omega_{\rm m} = 0.307$, $\Omega_{\rm b} = 0.04825$, $\sigma_8 = 0.8288$, $n_{\rm s} = 0.9611$ and $H_0 = 67.77 \rm km s^{-1} Mpc^{-1}$.

In order to predict the {\sc Sibelius} galaxy population, accounting for the environment and merger history, the semi-analytical galaxy formation model {\sc Galform} \cite{Cole-2000, Lacey-2016} was applied to its halo merger tree. In {\sc Galform}, galaxy properties such as the stellar mass and morphology are calculated based on the properties of the evolving dark matter structures, such as mass, accretion rate, spin, or mergers, with analytical prescriptions for processes including gas cooling, star formation, stellar evolution, black hole formation, feedback from supernovae and AGN, dynamical friction and dynamical instabilities. Mergers (particularly major mergers) and starbursts caused by disk instabilities are the primary processes that can lead to the growth of the bulge, and the transformation of a disk galaxy into an elliptical galaxy \cite{Cole-2000}.

Of particular relevance to this study is the fact that the  {\sc BORG} algorithm used in the construction of the constraints is agnostic to the astrophysical processes that may give rise to the divergent formation pathways of disks and ellipticals, while the {\sc Galform} model is agnostic to the particular structures present in the Local Universe, which, however, determine the properties of the merger trees. 

We classify the {\sc Sibelius} galaxies according to their bulge-to-disk mass ratio, $B/D$. We designate those with $B/D < 1/4$ as disk galaxies, those with $1/4 < B/D < 8$ as intermediates and those with $B/D > 8$ as ellipticals \cite{Benson-2010}, but our results are not sensitive to these particular choices. We focus here mostly on the distributions of disks and ellipticals, which are the subjects of \cite{Peebles-2022, Peebles-2022b}, but the intermediates provide a further point of comparison.

\begin{figure*}
    \includegraphics[height=0.735in]{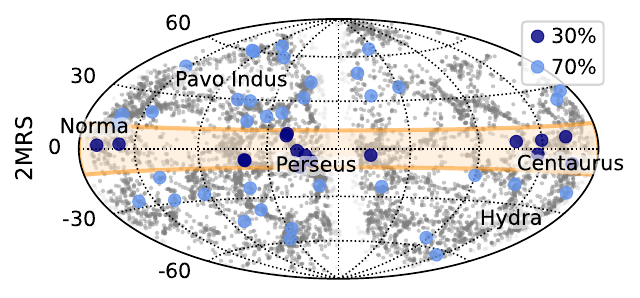}
    \includegraphics[height=0.735in]{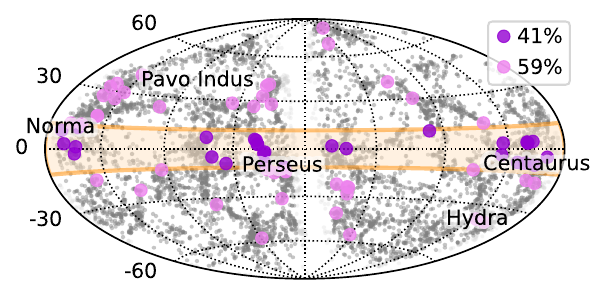}
    \includegraphics[height=0.735in]{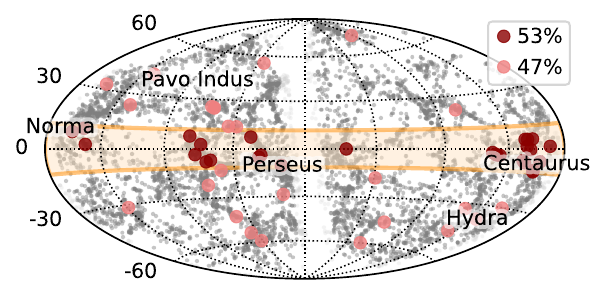}\\
        \vspace{-.2cm}
    \includegraphics[height=0.735in]{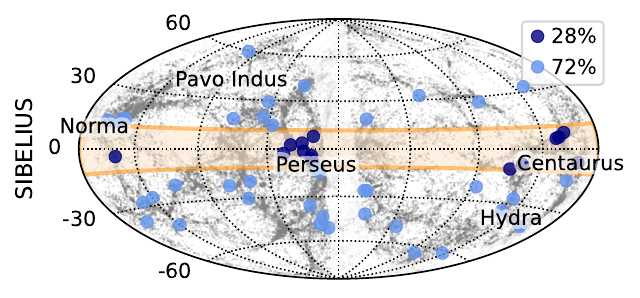}
    \includegraphics[height=0.735in]{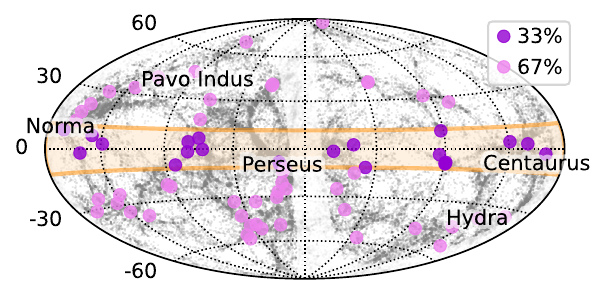}
    \includegraphics[height=0.735in]{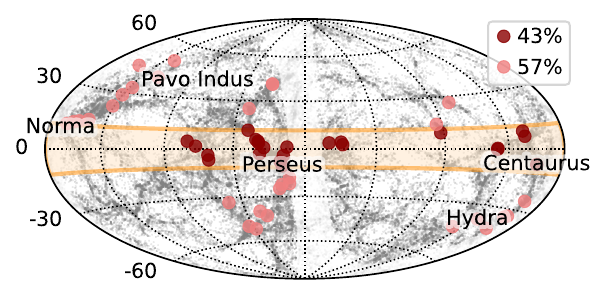} \\
    
    \caption{Distributions of disks, intermediates and ellipticals in the Local Universe up to $z = 0.02$. Hammer projection in supergalactic coordinates of the distribution of the most massive disks (left), intermediates (centre) and ellipticals (right) in the redshift range of $0.01 < z < 0.02$, in the 2MRS survey (top) and in {\sc Sibelius} (bottom). The shaded areas correspond to $\left\vert \mathrm{sin(SGB)}\right\vert  < 0.2$ $\left(\pm 11.5^\circ\right)$ around the supergalactic equator. Dark and light symbols show individual galaxies that lie inside and outside this region, respectively, and percentages in the top right of each panel express their relative numbers. Grey points denote lower mass galaxies (fainter for $\left\vert b\right\vert  < 10^\circ$) irrespective of morphology, labels indicate the positions of five galaxy clusters. In both the 2MRS data and the {\sc Sibelius} simulation, a significantly higher fraction of massive ellipticals than of massive disks lie close to the supergalactic plane.
    \label{fig:projection-bands}
    }
\end{figure*}

\subsection*{Predicted and observed distributions}
To compare to the observations, and in particular, to the results of \cite{Peebles-2022}, we select from the 2MASS galaxy redshift catalogue (2MRS, \cite{Huchra-2012}) the brightest objects in the redshift range $0.01 < z < 0.02$ by absolute $K$ band-magnitude. Excluding objects with absolute galactic latitude $\vert b \vert < 10^{\circ}$, with a magnitude limit of $M_K < - 25.19$ we obtain identical sample of galaxies composed of 54 disks, 53 ellipticals, and 73 intermediates.

In Fig.~\ref{fig:projection-bands}, we compare the distributions in supergalactic coordinates in 2MRS, at the top, and {\sc Sibelius-Dark}, at the bottom. From left to right, panels show the positions of the most massive $\mathrm{N}=54$ disks (in blue), $\mathrm{N}=73$ intermediates (in purple) and $\mathrm{N}=53$ ellipticals (in red). On all panels, grey dots show the positions of fainter galaxies tracing the local large-scale structure. Also shown are the locations of five galaxy clusters in this redshift range. Orange lines enclose a region of $\left\vert \mathrm{sin(SGB)}\right\vert < 0.2$ $\left(\mathrm{SGB} = \pm 11.5^\circ\right)$ around the supergalactic plane. Galaxies within and outside this region are denoted by dark and light symbols, respectively.

In both 2MRS and {\sc Sibelius}, the fraction of galaxies close to the supergalactic equator depends on their type. This fraction increases from disks to intermediates to ellipticals. The most massive ellipticals are more strongly clustered than the most massive disks. We explore the origin of this difference in the next sections.

\begin{figure*}
\centering
    \includegraphics[width=3.2in]{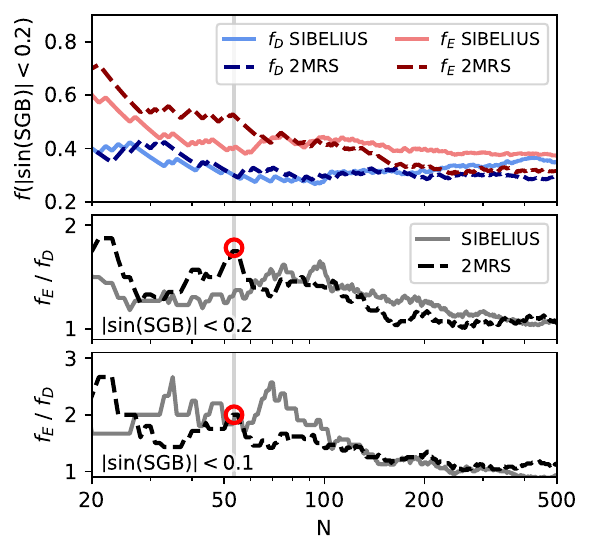}
    \caption{Excess of ellipticals relative to disks, as a function of sample size. Top: The fraction of the most massive N disks ($f_D$) and the most massive N ellipticals ($f_E$) in {\sc Sibelius} and 2MRS at $0.01 < z < 0.02$, located at $\left\vert \mathrm{sin(SGB)}\right\vert  < 0.2$. Centre: $f_E \ / f_D$, the fraction of the most massive N ellipticals divided by the fraction of the most massive N disks at $\left\vert \mathrm{sin(SGB)}\right\vert  < 0.2$. Bottom: ratio $f_E \ / f_D$, but now for $\left\vert \mathrm{sin(SGB)}\right\vert  < 0.1$. As indicated by the red circles, the values of $f_E / f_D$ in 2MRS peak at the value of N chosen in \protect\cite{Peebles-2022}. \label{fig:look-elsewhere}}
\end{figure*}

Due to the relatively small sample sizes, it is worth considering the possibility of a coincidence, or the `look elsewhere' effect. In Fig.~\ref{fig:look-elsewhere}, we compare the fractions of the most massive N disks, $f_D$, and the most massive N ellipticals, $f_E$, at $\left\vert \mathrm{sin(SGB)}\right\vert  < 0.2$ in 2MRS and {\sc Sibelius}, at $0.01 < z < 0.02$, as functions of N. Overall, we find good agreement between the simulation and the observations, and a significant excess, $f_E > f_D$, for all $\mathrm{N} < 150$ (Note that at large N, completeness may affect the results). We also find that in 2MRS, the relative excess peaks at the values used by \cite{Peebles-2022}. Comparisons at any other  $\mathrm{N}>25$ yield a smaller relative excess, indicating the presence of a 'look elsewhere' effect. In both {\sc Sibelius} and 2MRS, the relative excess increases for a narrower band of $\left\vert \mathrm{sin(SGB)}\right\vert  < 0.1$, albeit with even smaller sample sizes.

Drawing random sub-samples of 50 out of each of the 100 most massive disks and ellipticals in {\sc Sibelius} yields configurations where ellipticals are more clustered than disks (i.e. more objects within $\left\vert \mathrm{sin(SGB)}\right\vert  < 0.2$) in approximately 95.2\% of cases. This number decreases slightly to 93.5\% for $\left\vert \mathrm{sin(SGB)}\right\vert  < 0.25$, and to 84.5\% for $\left\vert \mathrm{sin(SGB)}\right\vert  < 0.15$. For 2MRS, the corresponding numbers are 88.0\%, 86.7\%, and 87.0\%. The excess is equally significant in {\sc Sibelius} and 2MRS.

\begin{figure*}
    \includegraphics[width=4.6in]{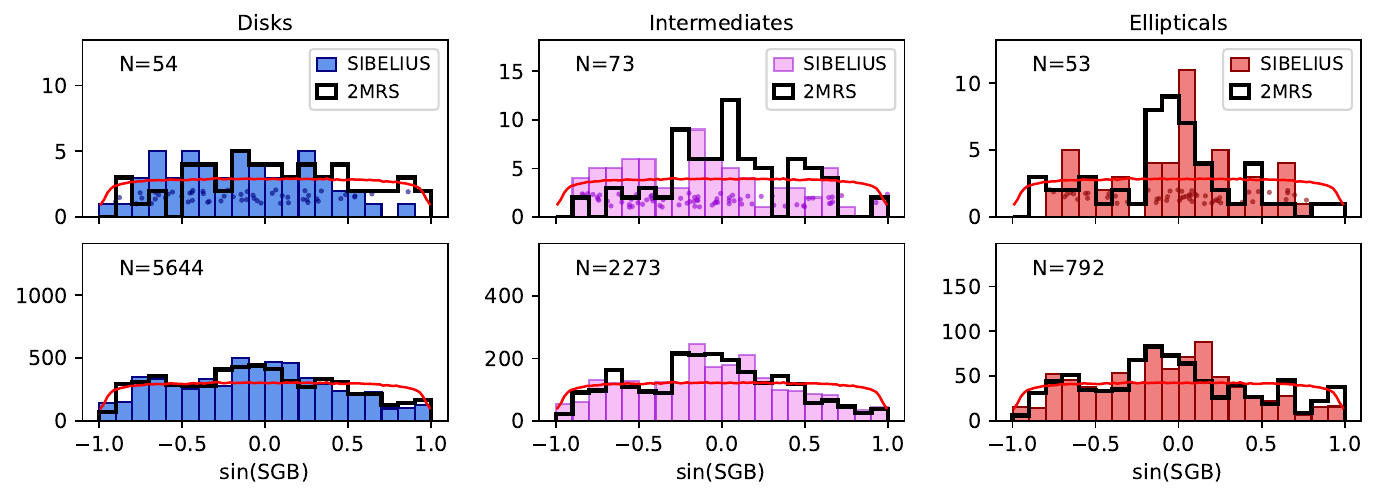}
    \caption{Distributions of supergalactic latitude (SGB) in the redshift range $0.01 < z < 0.02$ for galaxies of different morphology. Disks are shown on the left, intermediates in the centre, and ellipticals on the right. In each case, galaxies with $\vert b\vert < 10^\circ$ are excluded. Histograms show the 2MRS data and results from {\sc Sibelius}. The top row shows the N most massive objects in each class to match the samples of \protect\cite{Peebles-2022}, the bottom row shows the full 2MRS samples in this redshift range. In each case, we compare to an equal number of objects from {\sc Sibelius}. Red lines show the isotropic expectation accounting for sky coverage. In both the observations and simulation, the concentration towards the supergalactic equator is much more prominent in the most massive ellipticals than in the most massive disks; intermediates fall in between. By contrast, for the larger sample sizes, the distributions of different galaxy types are similar. \label{fig:histogram}}
\end{figure*}

In the top row of Fig.~\ref{fig:histogram}, we show the distributions of $\mathrm{sin\left(SGB\right)}$ at $0.01 < z < 0.02$ for the same numbers of most massive disks (left), intermediates (middle) and ellipticals (right) as used by \cite{Peebles-2022}. Black bars represent the 2MRS data, filled coloured bars represent {\sc Sibelius} for which points denote individual objects. As was already apparent from Fig.~\ref{fig:projection-bands} and Fig.~\ref{fig:look-elsewhere}, we find a strong excess towards low supergalactic latitudes of the most massive ellipticals in {\sc Sibelius}, in good agreement with 2MRS. There is a much weaker enhancement for the most massive disks, also in good agreement with observations. In both cases, the distributions of the most massive intermediates fall in between: they are more strongly clustered than the most massive disks, but less so than the most massive ellipticals.

When we expand our samples to match the respective sample sizes of the 2MRS catalogue, as shown in the bottom row of Fig.~\ref{fig:histogram}, we find broadly similar distributions for disks, intermediates, and ellipticals, and excellent agreement between the simulation and the observations. To assess the significance of the agreement of the SGB distributions in {\sc Sibelius} and 2MRS, we generate random permutations of the observational data. This gives a probability for the conspicuous agreement in the bottom row of Fig.~\ref{fig:histogram} to arise by chance of $< 0.001 \%$ for either disks or intermediates, and $\sim 0.02 \%$ for ellipticals (which have a smaller sample size).

The results of {\sc Sibelius} thus reproduce those presented in \cite{Peebles-2022} derived from 2MRS: there is a clear difference in clustering between ellipticals and disks with respect to the supergalactic plane, but this is only manifest among the most massive objects, not in the much larger samples.

\begin{figure*}
   \includegraphics[width=1.53in]{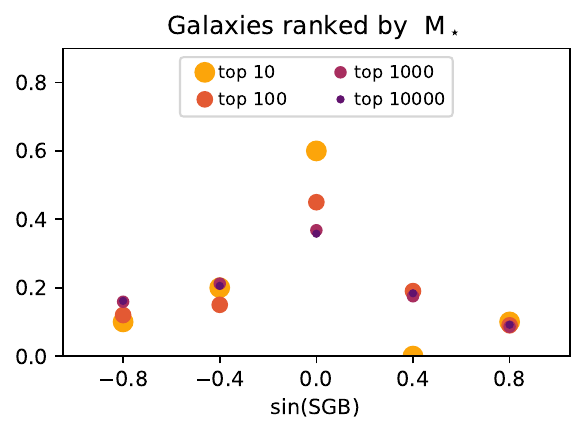}
    \includegraphics[width=1.53in]{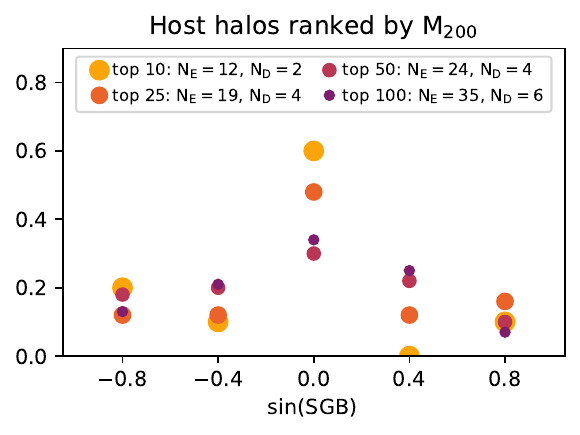}
    \includegraphics[width=1.53in]{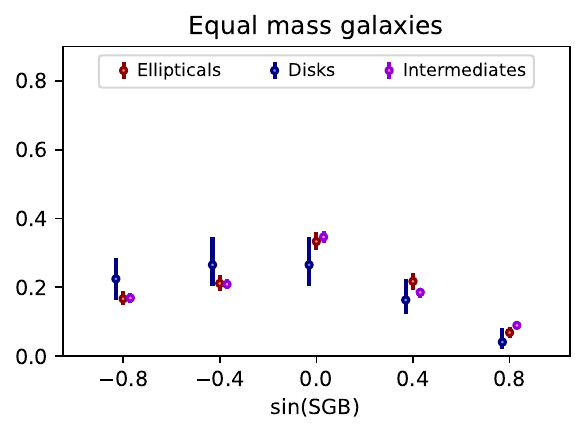}
    \caption{Distributions of supergalactic latitude (SGB) in the redshift range $0.01 < z < 0.02$ for simulated galaxies and host haloes ranked according to different criteria.
    {\it Left:} distributions of galaxies ranked by stellar mass the 10 or 100 most massive galaxies in the {\sc Sibelius} simulation are significantly more strongly concentrated to low supergalactic latitudes compared to less massive galaxies. The distributions of the 1000 or 10000 most massive galaxies are similar. Centre: distribution of the most massive haloes, ranked by $\mathrm{M_{200}}$. The most massive haloes (which contain a high number, $\mathrm{N_E}$, of the 50 most massive ellipticals, but few, $\mathrm{N_D}$, of the 50 most massive disks) are close to the supergalactic plane. {\it Right:} distributions of ellipticals, disks and intermediates in the mass range of the most massive 100 disks, i.e. at the same stellar mass. Bars indicate $\pm 1 \sigma$ statistical error. Even comparing at equal mass, ellipticals and intermediates are more strongly concentrated near the supergalactic plane than disk galaxies.
    \label{fig:histogram-analysis}
    }
\end{figure*}

\subsection*{Causes of the separation}\label{sec:mass}
We identify several causes for the different distributions of the most massive disks and ellipticals in the {\sc Sibelius} simulation, and by implication, in the Local Universe. We first note that most massive ellipticals in {\sc Sibelius} are much more massive than the most massive disks, and the most massive galaxies in {\sc Sibelius} are overwhelmingly elliptical. For instance, while the 100 most massive disks in the redshift range $0.01 < z < 0.02$ have a median stellar mass of $2.9 \times 10^{10}~\Ms$ (s.d. $ 0.8 \times 10^{10}~\Ms$), the corresponding figure for the 100 most massive ellipticals is $3.0 \times 10^{11}~\Ms$ (s.d. $1.7 \times 10^{11}~\Ms$), an order of magnitude higher.

In the left panel of Fig.~\ref{fig:histogram-analysis}, we compare the distributions of $\mathrm{sin\left(SGB\right)}$ for galaxies of different stellar mass, irrespective of galaxy type. The most massive 10 or 100 galaxies, with stellar masses above $5 \times 10^{11}~\Ms$ and $2.5 \times 10^{10}~\Ms$, respectively, are significantly more concentrated towards low SGB than less massive galaxies.

This mass difference is in agreement with observational studies of galaxies in the Local Universe \cite{Bernardi-2003, Vulcani-2011, Ryan-2012}, which also find that the most massive galaxies are overwhelmingly elliptical. Using HI and UV data, \cite{Jackson-2022} recently measured the fraction of ellipticals among galaxies with stellar masses above $10^{11.4} \ \Ms$ to be $\sim 87\%$. Hydrodynamic simulations \cite{Jackson-2020} and the {\sc Galform} semi-analytical model, both of which are calibrated to reproduce the observed mass functions qualitatively reproduce these findings. This suggests that the mass difference we identify in {\sc Sibelius} reflects that in the real Universe, and explains part of the difference in the distributions of $\mathrm{sin\left(SGB\right)}$, in line with the standard model prediction \cite{Kaiser-1987} and observations \cite{Wake-2011}, that more massive galaxies are more strongly clustered.

It is worth noting that the $K$-band magnitudes in 2MASS for the most massive disks and ellipticals are in fact similar \cite{Peebles-2022}. We attribute this to the previously reported systematic underestimation of the magnitudes of massive ellipticals in 2MASS \cite{Lauer-2007, Schombert-2012, Kormendy-2013, Laesker-2014, Ma-2014}. Applying the correction suggested by \cite{Laesker-2014} to the ellipticals would result in a much larger fraction of ellipticals among the brightest galaxies, in line with the other studies mentioned above and with {\sc Sibelius}. Of course, results based on ranking galaxies within the same morphological class remain unchanged.

We also find evidence that the different environments on or off the supergalactic plane act to separate the most massive galaxies by morphology, in addition to simply sorting them by mass. In the centre panel of Fig.~\ref{fig:histogram-analysis}, we show the distributions in $\mathrm{sin\left(SGB\right)}$ of host haloes in different mass ranges. As expected, the most massive 10 or 25 haloes in {\sc Sibelius} in this redshift range are strongly biased towards the supergalactic plane. They also host a significant number, $\mathrm{N_E}$, of the 50 most massive ellipticals, but only a small number, $\mathrm{N_D}$, of the 50 most massive disks. For example, 19 of the 50 most massive ellipticals, but only 4 of the 50 most massive disks, are in the 25 most massive haloes. The 50 most massive ellipticals reside in 45 haloes with a median mass of $M_{200} = 9.0 \times 10^{13}~\Ms$, while the 50 most massive disks are all in different haloes with a median mass of $M_{200} = 1.9 \times 10^{12}~\Ms$, over 50 times lower.

In the right panel of Fig.~\ref{fig:histogram-analysis}, we show the distribution of galaxies of different types, ellipticals, disks, and intermediates, in the mass range of the 100 most massive disks. All three galaxy samples have very similar masses, but the disk galaxies are still significantly less strongly clustered than ellipticals or intermediates of similar mass. This further underlines the role of the environment: galaxies of the same mass are less likely to evolve into massive disks in the dense environment that marks out the supergalactic plane.

In the Horizon-AGN hydrodynamical simulation, two scenarios for the formation of massive disk galaxies are identified \cite{Jackson-2020}: an exceptionally quiet merger history, or a late merger of a spheroid with a massive, gas-rich satellite. Both of these are less likely to occur in a dense environment, where mergers are more common, and cold gas fractions are lower. This provides an additional, environmental  explanation for the relative paucity of massive disks near the supergalactic plane.

\begin{figure*}
    \includegraphics[height=0.85in]{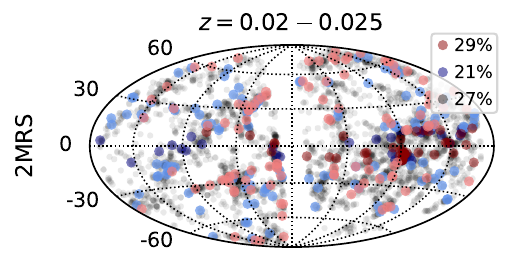}        \includegraphics[height=0.85in]{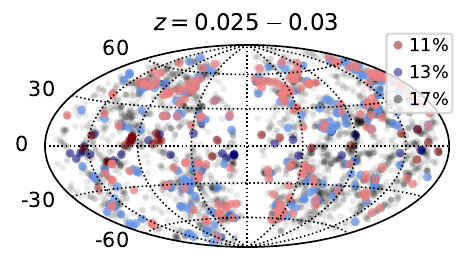}  
    \includegraphics[height=0.85in]{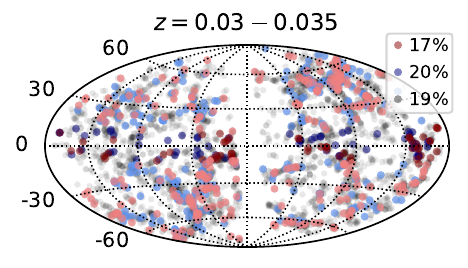}  \\
        \vspace{-.1cm}
    \includegraphics[trim=0 0 0 0.2cm, clip, height=0.75in]{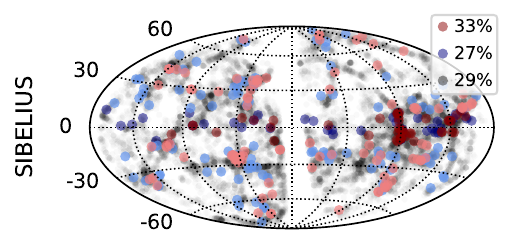}           \includegraphics[trim=0 0 0 0.2cm, clip, height=0.75in]{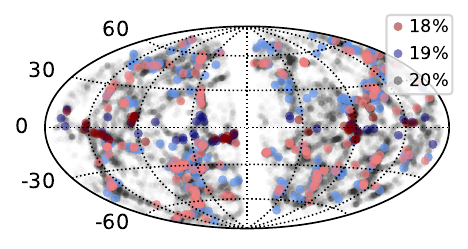}  
    \includegraphics[trim=0 0 0 0.2cm, clip, height=0.75in]{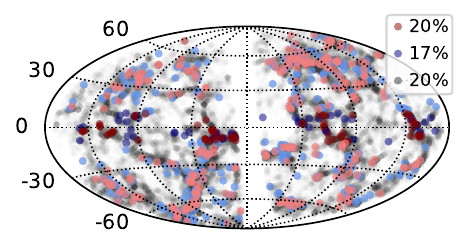}  \\

    \caption{Distributions of galaxies beyond $z=0.02$. Hammer projection in supergalactic coordinates of the distributions of the most massive disks and ellipticals in 2MRS (top) and {\sc Sibelius} (bottom) in three redshift slices. Galaxies are selected at $\vert b\vert  > 10^\circ$ and with masses similar to the 100 most massive galaxies of each type at $z=0.01-0.02$. As in Fig.~\ref{fig:projection-bands}, dark and light circles denote galaxies within and outside of $\pm 11.5^\circ$ of the supergalactic plane, respectively. Fainter galaxies are shown in grey. Percentages indicate the fraction of galaxies in each class that are within  $\pm 11.5^\circ$ of the supergalactic plane. The ellipticals more closely trace the distribution of the most prominent structures, while the disks are more equally distributed. The excess of galaxies at low supergalactic latitudes extends to $z\sim 0.025$. At higher redshifts, the distributions of both structures and galaxies are no longer aligned with the supergalactic plane.
    \label{fig:projection-outside}
    }
\end{figure*}

\subsection*{The extent of the supergalactic plane}
That {\sc Sibelius} is designed to reproduce structures out to $\sim 200$ Mpc also lets us investigate the extent of the supergalactic plane beyond $z=0.02$. Fig.~\ref{fig:projection-outside} shows the sky distributions of disks and ellipticals in 2MRS (top) and {\sc Sibelius} (bottom) in redshift shells out to  $z=0.035$. 

The overdensity at low supergalactic latitudes only extends to $z \sim 0.025$. Beyond this, structures no longer align with the supergalactic plane, the distributions become less dominated by individual clusters, and the fraction of galaxies near the equator fluctuates around the isotropic expectation value of $\sim 20\%$. However, in every redshift range, both in 2MRS and {\sc Sibelius}, elliptical galaxies are more strongly clustered than disks. We also investigated the presence of assembly bias, i.e. the difference in clustering that is not explained by halo mass \cite{Croton-2007}. However, for the comparatively small galaxy samples we consider here, we found no significant effect.

\begin{figure*}
\centering
    \vspace{-.1cm}
    \includegraphics[width=2.3in]{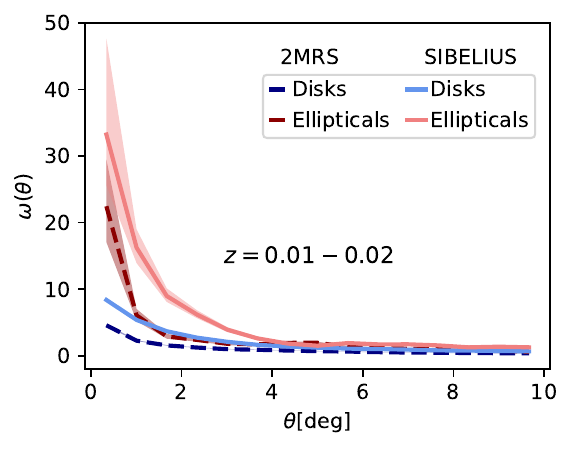} 
    \includegraphics[width=2.3in]{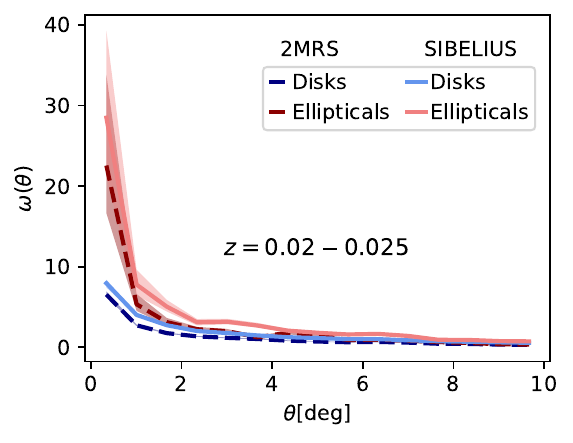} \\
    \includegraphics[width=2.3in]{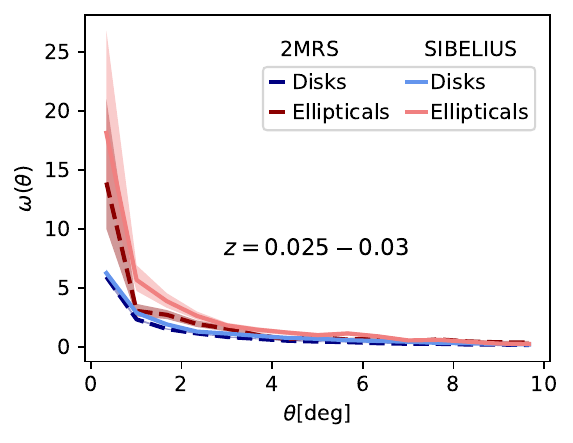}
    \includegraphics[width=2.3in]{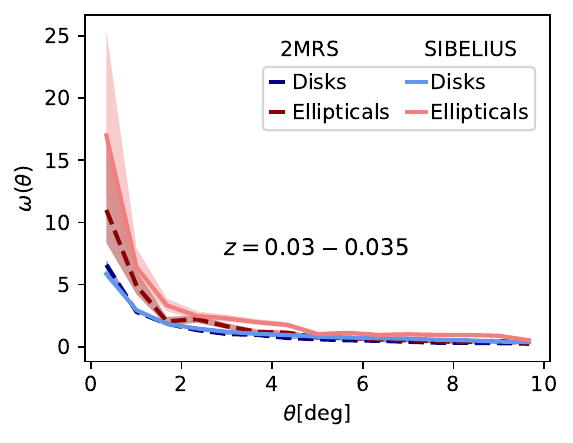} \\

    \caption{Angular correlation functions. Lines show $\omega (\theta)$ of disks (blue) and ellipticals (red) in the {\sc Sibelius} simulation (light solid) and the 2MRS data (dark dashed lines) in four redshift ranges. The top left panel corresponds Fig.~\ref{fig:projection-bands} while the top right panel and bottom row corresponds to Fig.~\ref{fig:projection-outside}. Shaded areas provide an estimate of the statistical uncertainty. In each redshift range, within and beyond the Local Supercluster, and in both 2MRS and {\sc Sibelius}, elliptical galaxies show significantly stronger clustering than disks.
    \label{fig:correlation}
    }
\end{figure*}

The enhanced clustering of ellipticals compared to disks also manifests in the angular correlation functions shown in Fig.~\ref{fig:correlation}. In both the simulation and the 2MRS data, and in every redshift range, ellipticals are significantly more clustered than the disks. The peculiar excess of massive ellipticals near the supergalactic plane at redshifts $0.01 < z < 0.02$ is not caused by exceptional galaxy formation physics but simply reflects the particular structures in this region.

\section*{Discussion}\label{sec:conclusion}
Constrained simulations such as {\sc Sibelius} provide a unique window onto the local matter and galaxy distributions, and allow powerful, direct tests of cosmological and astrophysical models. While environmental effects on morphology have been previously demonstrated \cite{Zehavi-2005, Coil-2008, Farrow-2015, Hatfield-2019, Smith-2022}, earlier works based on random environments rely on the assumption that cosmic variance will cancel out. By comparing theory and observations in matched environments, we do not have to make this assumption.

The {\sc Sibelius} simulation, in combination with the {\sc Galform} semi-analytical model, reproduces the distributions of ellipticals, intermediates and disks in the redshift range $0.01 < z < 0.02$ and beyond. Importantly, it also reproduces the striking difference in the clustering of the most massive ellipticals and disks vis-a-vis the supergalactic plane.

We identify two causes for this dichotomy in {\sc Sibelius}. Firstly, the most massive ellipticals in the redshift range $0.01 < z < 0.02$ are much more massive than the most massive disks, and the clustering strength depends strongly on galaxy mass. Galaxies which fall between the ellipticals and disks in terms of both morphology and stellar mass, also follow an intermediate spatial distribution. Secondly, from the fact that the most massive disks are less strongly clustered than intermediates or ellipticals {\it of the same mass}, we conclude that the environment prevailing in the supergalactic plane inhibits the conditions necessary for massive disk formation: a quiet merger history and the continuous supply of cold gas. In {\sc Sibelius}, a large fraction of the most massive ellipticals are close to the supergalactic plane by virtue of residing within the galaxy clusters that define it. 

The {\sc Galform} galaxy formation model is idealised and cannot capture the full complexity of astrophysical processes at play in hydrodynamic simulations, let alone in the real universe. Nevertheless, it is patently sufficient to reproduce the distinct populations of elliptical and disk galaxies observed in the Local Universe.

The strikingly different distributions of bright ellipticals and disks vis-a-vis the supergalactic plane do not require physics beyond the standard model. They arise naturally in the $\Lambda$CDM framework and the standard model of galaxy formation represented by the {\sc Sibelius} simulation of the Local Universe. They are a direct consequence of the differences in mass and in the associated bias, and of the transformation of disks into ellipticals within the clusters that define the supergalactic plane. Rather than as an anomaly, the observed distributions emerge as a prediction of the $\Lambda$CDM paradigm, and as an important benchmark for any alternatives.

\newpage

\section*{Methods}

\subsection*{The {\sc Sibelius Dark} simulation}
Our work is based on the {\sc Sibelius Dark} simulation ({\sc Sibelius} for short throughout this paper), a collisionless cosmological simulation which embeds a high-resolution constrained region of radius $\sim 200$~Mpc centred on the Local Group within a 1~Gpc$^3$ periodic volume. {\sc Sibelius} is set up with cosmological parameters $\Omega_\Lambda = 0.693$, $\Omega_m = 0.307, \Omega_b = 0.042825$, $\sigma_8 = 0.8288$, $n_s = 0.9611$ and $H_0 = 67.77~\rm km s^{-1}~Mpc^{-1}$. The high resolution region is sampled with $\sim 130$ billion dark matter particles of mass $1.15 \times 10^7 \Ms$ with a maximum physical softening length of $1.4$ kpc. More details of the numerical setup are given in \cite{McAlpine-2022}, which also includes additional comparisons to observations. 

The initial conditions for {\sc Sibelius} are created using white noise fields obtained through the {\sc BORG} algorithm \cite{Jasche-2013, Lavaux-2016, Jasche-2019}, which takes as an input the observed three-dimensional density field, inferred in this instance from the 2M++ galaxy survey \cite{Lavaux-2011}. {\sc BORG} is designed to infer the most likely initial configuration that gave rise to the input galaxy density distribution, subject to the specified cosmology and an assumed galaxy bias model. As such, it is designed to represent the most likely underlying matter distribution and formation history of the structures in the Local Universe. It is important to note, however, that the input galaxy distribution used in constructing the initial conditions for {\sc Sibelius} does not differentiate by galaxy morphology or colour. The reproduction of the observed local large-scale structure is thus {\it by construction}, but the reproduction of distinct galaxy sub-populations is not.

Outputs of {\sc Sibelius} are stored at 200 times between $z=25$ and $z=0$. For each of these snapshot, haloes and subhaloes are identified using a Friend-of-Friend algorithm and the Hierarchical-Bound-Tracing (HBT) algorithm \cite{Han-2012}, respectively. Subhaloes are further processed into a merger tree, to which the {\sc Galform} semi-analytical model is applied, as discussed below.

\subsection*{The {\sc Galform} semi-analytical model}
{\sc Sibelius Dark} only traces the evolution of matter via gravity. To derive a prediction for the galaxy population that can be compared to observations, the separate evolution of baryons and the formation of galaxies is followed using the semi-analytical model {\sc Galform}. All semi-analytical models make the same basic assumptions: that baryons follow the accretion of matter (including its angular momentum) on to collapsed dark matter haloes \cite{White-1978}, and that the evolution of baryons inside dark matter haloes and during mergers and interactions can be described by a set of coupled, non-linear differential equations \cite{White-1991, Cole-1994, Baugh-2006}. In {\sc Galform}, these equations describe and account for various astrophysical processes including gas cooling, dynamical friction, star formation, stellar evolution, black hole formation and evolution, and the formation and feedback from AGN, stars and supernovae.

The parameters of the model are set either from ab initio considerations, or via calibration to observations of the galaxy population. Since its first introduction \cite{Cole-2000} and through several subsequent updates \cite{Benson-2000, Bower-2006, Lacey-2016}, when applied to $\Lambda$CDM merger trees, {\sc Galform} has been shown to reproduce many of the observed properties of the galaxy population, such as the HI or stellar mass functions, the ratio of ellipticals to disks, and the Tully-Fisher relation. It has also been shown to reproduce the galaxy population as well as the properties of individual galaxies (albeit with considerable scatter) in the {\sc EAGLE} hydrodynamic simulation \cite{Mitchell-2018}. The version of {\sc Galform} used for {\sc Sibelius Dark} \cite{Lacey-2016} incorporates different initial stellar mass functions during starbursts, feedback that suppresses gas cooling in massive haloes, and a new empirical star formation law in galaxy discs based on molecular gas content. While it still lacks some of the complexity of a full hydrodynamic simulation, {\sc Galform} is designed to implement our current understanding of galaxy formation physics. 

Importantly, the calibration of its free parameters has been performed using merger trees from randomised initial conditions without consideration of galaxy clustering \cite{Lacey-2016} and independently of the constraints used in {\sc Sibelius}, or of the particular structures present in the Local Universe.

\setcounter{figure}{0}
\renewcommand{\figurename}{Extended Data Fig.}

\begin{figure*} 
    \centering
    \includegraphics[width=3.in]{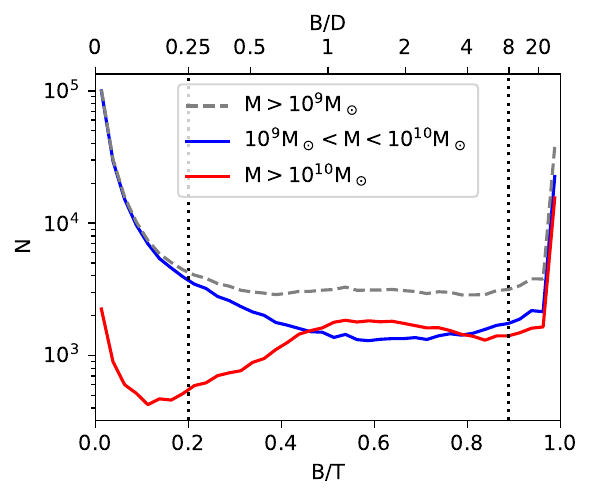} 
    \caption{Distribution of galaxy morphologies as parameterised by $B/D$ or $B/T$ in the {\sc Sibelius} simulation in the redshift range $z=0.01 - 0.04$. The blue curve shows galaxies in the stellar mass range $10^9-10^{10} \Ms$, the red curve shows galaxies above $10^{10} \Ms$, the grey dashed curve shows all galaxies above $10^9 \Ms$. There is a clear bimodality in the bulge fraction, and the fraction of galaxies with high bulge fraction increases with increasing stellar mass. The two vertical dotted lines denote $B/D = 1/4 \ (B/T = 1/5)$, our default threshold separating  disks and intermediates, and $B/D = 8 \ (B/T = 8/9)$, our default threshold separating intermediates and ellipticals.
    \label{fig:morphologies}}
\end{figure*}

\subsubsection*{Separation of disks, ellipticals and intermediates}
As shown in Extended Data Fig.~\ref{fig:morphologies}, the {\sc Galform} semi-analytical model results in a bimodal distribution of galaxy morphologies as characterised by the bulge-to-disk mass ratio, $B/D$, or equivalently, bulge-to-total mass ratio, $B/T$. We use $B/D$ to classify {\sc Sibelius} galaxies into disks, ellipticals and intermediates. By default, we designate galaxies with $B/D < 1/4$ as disks, those with $1/4 < B/D < 8$ as intermediates, and those with $B/D > 8$ as ellipticals \cite{Benson-2010}. These two limits are indicated by dotted vertical lines in Extended Data Fig.~\ref{fig:morphologies}. Using these values, $90\%$ of disks have $B/D < 0.13$, while $90\%$ of ellipticals have $B/D > 12.6$, indicative of a strongly bimodal distribution. It can also be seen that the morphology distribution depends on mass. In agreement with observations~\cite{Benson-2007, Gonzalez-2009}, the lower mass galaxy population contains a larger fraction of disks, while higher mass galaxies are more likely to be elliptical.

\begin{figure*} 
\centering
    \includegraphics[width=2.3in]{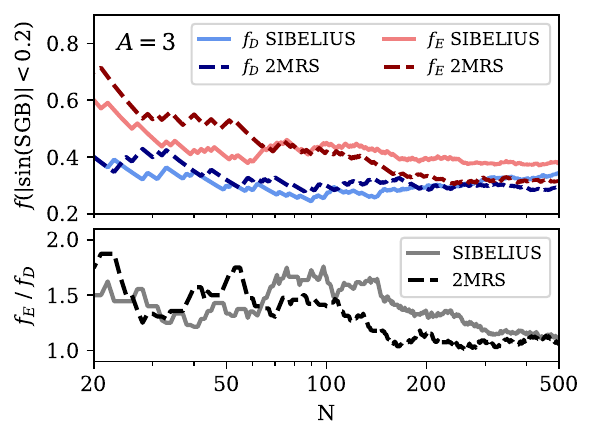} \includegraphics[width=2.3in]{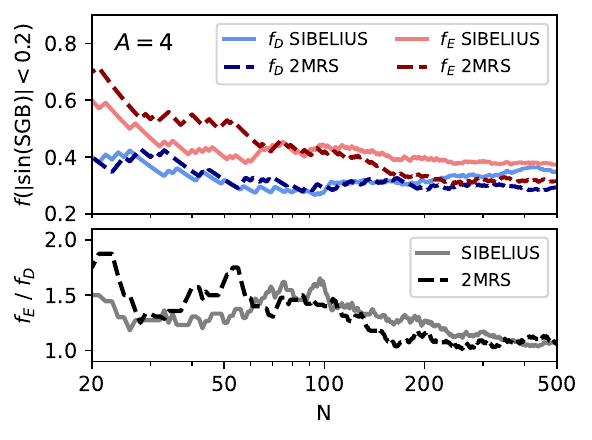} \\
    \includegraphics[width=2.3in]{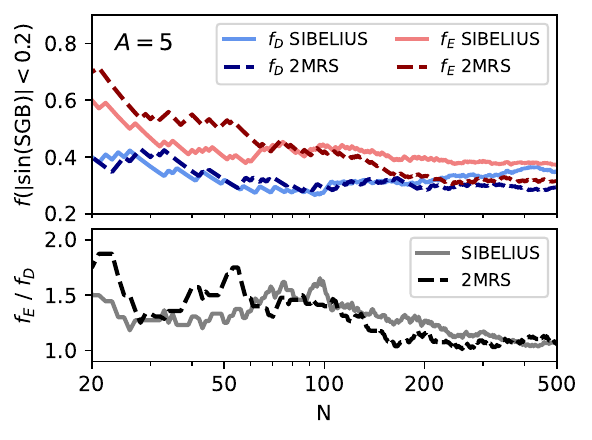} 
    \includegraphics[width=2.3in]{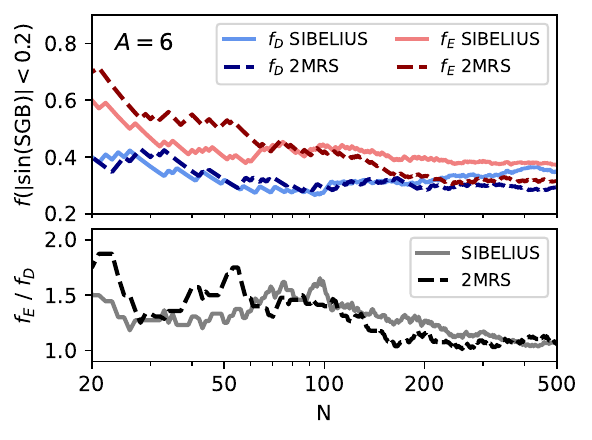}
    \caption{The effect of adopting different thresholds to delineate disks, intermediates and ellipticals in {\sc Sibelius}. Analogous to Fig.~\ref{fig:look-elsewhere}, each panel shows the excess of ellipticals and disks at $\left\vert \mathrm{sin(SGB)}\right\vert < 0.2$, for different delineations between disks and ellipticals in {\sc Sibelius}, as parameterised by bulge-to-disk ratios, less than $1/A$ for disks and greater than $A$ for ellipticals. The top row shows, from left to right, $A = 3$ and $A = 4$, while the bottom row shows $A = 5$ and $A = 6$. For reference, the 2MRS data is repeated on each panel. The greater abundance of ellipticals compared to disks at low supergalactic latitudes in {\sc Sibelius}, and its dependence of $N$, is not sensitive to the particular choice of $A$.
    \label{fig:look-elsewhere-A}}
\end{figure*}

Because the thresholds separating the three classes are still somewhat arbitrary, we also investigate the effect of choosing different values. In Extended Data Fig.~\ref{fig:look-elsewhere-A}, we repeat Fig.~\ref{fig:look-elsewhere} using $B/D < 1/A$ for disks and $B/D > A$ for ellipticals, for values of $A = 3, 4, 5$ and 6. It can be seen that the behaviour, in terms of an excess clustering of the most massive ellipticals relative to disks, is similar in every case, demonstrating that our results are not sensitive to the precise delineation between galaxy types.

\subsection*{The 2MASS redshift survey (2MRS) }
The observed distributions of galaxies are drawn from the 2MASS (Two Micron All Sky Survey) Redshift Survey (2MRS) \cite{Huchra-2012}. In total, the 2MRS catalogue contains 44 599 galaxies. We include in our analysis all galaxies for which redshift and morphological types are available. Galaxy morphologies are as assigned in \cite{Huchra-2012},
and we follow Peebles' classification, whereby types $ \geq 1$ are classified as ``spirals" or ``disks", and types $\leq -5$ are classified as ``ellipticals". We additionally identify intermediate types as ``intermediates".

To be able to sort galaxies by luminosity, we infer absolute K-band magnitudes using the distance modulus. We adopt $H_0 = 70$ kms$^{-1}$Mpc$^{-1}$ for consistency with \cite{Peebles-2022}, but we obtain the same galaxy samples with a value of $H_0 = 67.77$ kms$^{-1}$Mpc$^{-1}$ as used in the {\sc Sibelius} simulation. We also adopt the same galactic latitude threshold $\vert b \vert > 10^\circ$ to obtain from the 180 brightest galaxies in the redshift range $0.01 < z < 0.02$ the same samples of 54 brightest disks and 53 brightest ellipticals. We classify the remaining 73 as intermediates. We use the same morphology definitions and galactic latitude thresholds also at higher redshifts, up to $z \sim 0.035$, and we apply the same galactic latitude cuts to the simulation data.

To perform the comparison in redshift space, we use the measured redshifts for the 2MRS sample, and use a value of $H_0 = 67.77~\rm km s^{-1}~Mpc^{-1}$ (in addition to the proper motions) when assigning redshifts to the galaxies in {\sc Sibelius} based on their comoving simulation coordinates.

\subsubsection*{Magnitude bias for ellipticals}
As noted by Peebles, the absolute K-band magnitudes of the brightest 53 disk and brightest 54 elliptical galaxies at $0.01 < z < 0.02$ in 2MRS are similar, which would indicate similar stellar masses \cite{Kauffmann-1998}. This is in contrast to other observations, to the results of hydrodynamic simulations, and also to the results of the {\sc Galform} semi-analytical model, according to which the most massive ellipticals are significantly more massive than the most massive disks, and by implication, the most massive galaxies are overwhelmingly elliptical. We attribute this difference to the known result that 2MASS systematically and significantly underestimates the luminosity of bright bulges and ellipticals \cite{Lauer-2007, Schombert-2012, Kormendy-2013, Laesker-2014, Ma-2014}. 

In particular, by comparison to deeper K-band observations of the same objects, \cite{Laesker-2014} found that 2MASS underestimates the luminosity of ellipticals and bulges by an average of 0.34 magnitudes, with the effect increasing for bright ellipticals. Applying a (somewhat conservative, but also crude) correction of $-0.34$ to the magnitudes of all ellipticals changes the distribution of morphological types among the brightest galaxies. Using the same selection as \cite{Peebles-2022}, among the 180 brightest galaxies in 2MRS with  $0.01 < z < 0.02$ and $\vert b \vert > 10^\circ$, the number of ellipticals increases from 53 to 89, while the number of disks decreases from 54 to 34. Among the brightest 100 galaxies, the number of ellipticals increases from 36 to 56, while the number of disks decreases from 25 to 14.

It is important to note that this correction would not change the ordering of galaxies within the same class, the identity and positions of the brightest N ellipticals or the brightest N disks remain the same. As such, we have no need to apply a magnitude correction in this work, but its effect may explain why, when the 2MASS K-band magnitudes were taken at face value, stellar mass and the associated formation bias was excluded out as a factor for explaining the different distributions of bright ellipticals and disks \cite{Peebles-2022}.

\subsection*{Angular correlation functions}
As noted in the main text, the strikingly different distributions of bright ellipticals and disks vis-a-vis the supergalactic plane reflects the fact that, in the redshift range $0.01 < z < 0.02$, the Local Supercluster is the most prominent feature, and aligns with the supergalactic equator. In other redshift ranges, we find no strong excess of ellipticals near the supergalactic plane, but as shown in Fig.~\ref{fig:correlation}, the difference between the correlation functions of disks and ellipticals persists at all redshifts, both in 2MRS and in {\sc Sibelius}. The astrophysical process that lead to different clustering of disks and ellipticals are universal, but the prominent excess of ellipticals at low supergalactic latitudes in the redshift range $0.01 < z < 0.02$ is caused by the presence of the Local Supercluster in this region of the Local Universe.

To compute the angular correlation functions, $\omega(\theta)$, we use the Landy-Szalay estimator \cite{Landy-1993}. For each redshift range and each galaxy type, we use the full 2MRS sample, choosing matching numbers of most massive galaxies in {\sc Sibelius} and of random points. We apply the same $\vert b \vert > 10^\circ$ mask for each data set and for the random samples. We use the jackknife method to account for the random nature of galaxy formation in every data set and also repeat the random sampling to estimate the statistical uncertainty.

\begin{figure*}
\centering
    \includegraphics[height=1.19in]{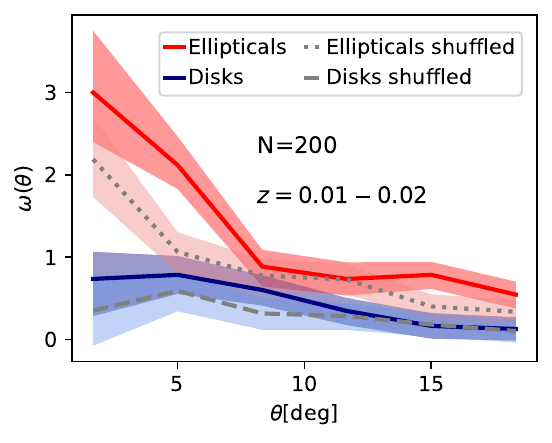} 
    \includegraphics[height=1.19in]{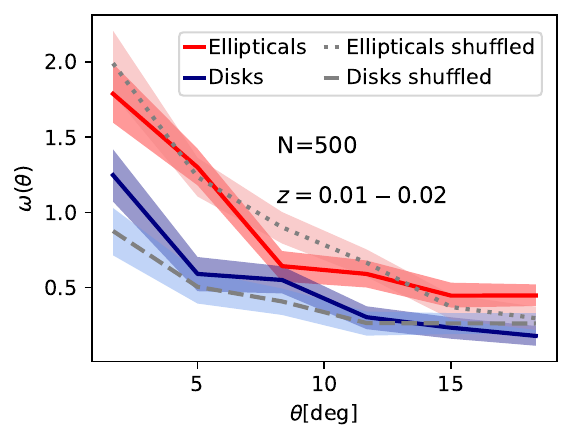} 
    \includegraphics[height=1.19in]{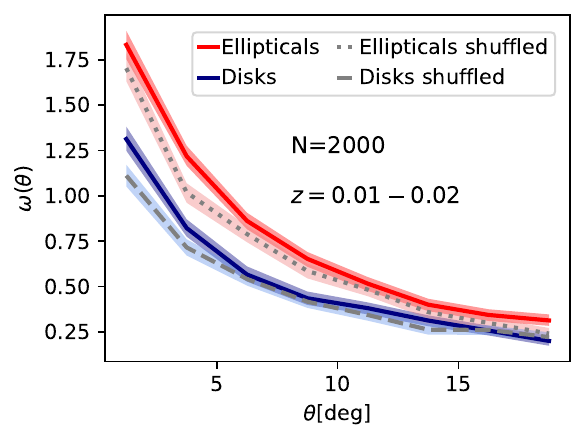} \\
    \includegraphics[height=1.19in]{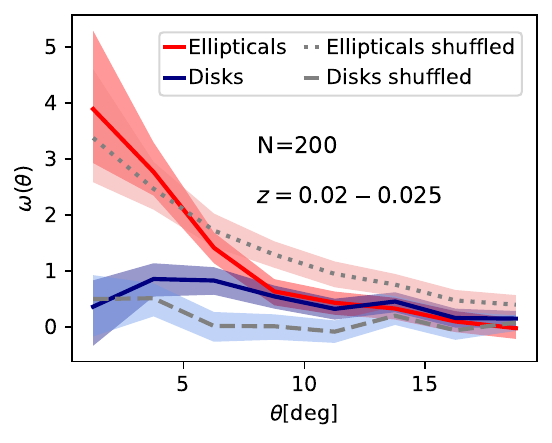} 
    \includegraphics[height=1.19in]{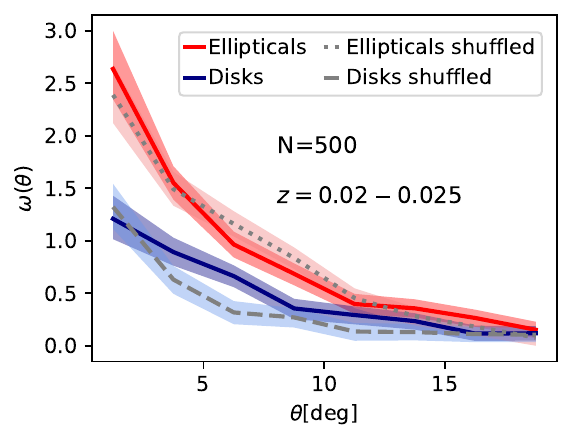} 
    \includegraphics[height=1.19in]{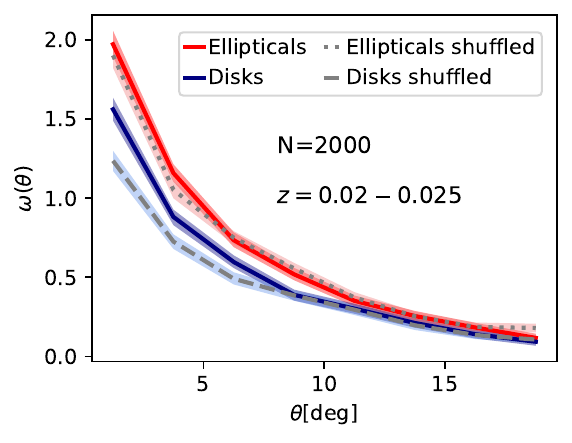} \\
    \includegraphics[height=1.19in]{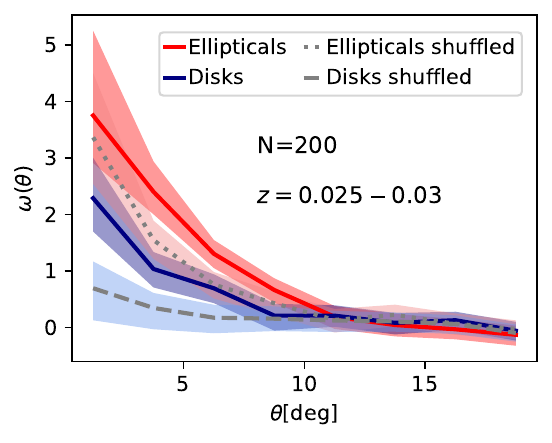} 
    \includegraphics[height=1.19in]{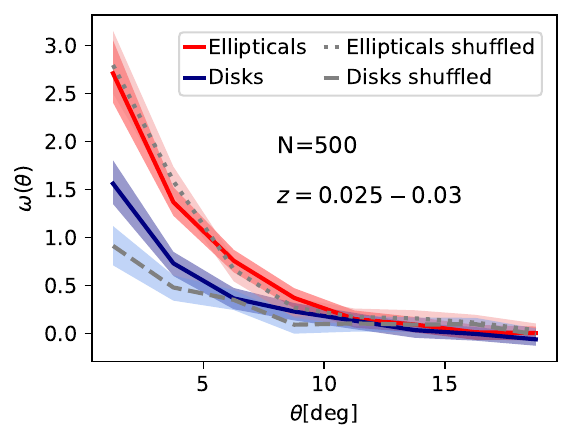} 
    \includegraphics[height=1.19in]{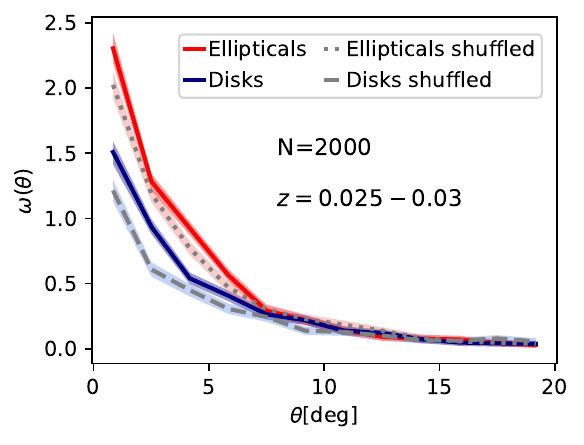} \\
 
    \caption{Assembly Bias. Angular correlation functions for samples of the $N$ most massive central ellipticals and central disks in three redshift intervals, and for the corresponding "shuffled" samples wherein each galaxy is replaced by another central galaxy of similar host halo mass regardless of galaxy type. Note that the correlations are weaker than those in Fig.~\ref{fig:correlation} due to the absence of the one-halo term. Assembly bias manifests itself as a difference between the original and shuffled samples, but we detect no clear effect of assembly bias for these sample sizes.
    \label{fig:assembly}
    }
\end{figure*}

\subsection*{Assembly bias}
Elliptical galaxies populate much more massive haloes than disk galaxies in our simulation. As is evident from Fig.~\ref{fig:histogram} and Fig.~\ref{fig:histogram-analysis}, respectively, the stronger clustering of ellipticals towards the supergalactic plane at $z=0.01-0.02$ is present both when classifying galaxies by morphology or by host halo mass.

This may raise the question of whether the stronger clustering of ellipticals is driven entirely by the mass of their host haloes, or whether there is an additional effect, known as assembly bias~\cite{Croton-2007}. Assembly bias can be measured by comparing the correlation functions of samples of galaxies separated by morphology to those obtained when the position of each galaxy in the sample is randomly assigned to a halo of similar mass, regardless of morphology. A systematic difference between the correlation function of the original sample and that of the shuffled counterpart must be attributable to factors other than halo mass, such as environment or assembly history, and is referred to as assembly bias. \cite{Croton-2007} found that assembly bias accounts for approximately $10\%$ of the total bias, being positive for ellipticals, slightly negative for disks, and decreasing in magnitude for the most massive galaxies.

We performed a similar test for the galaxy populations in {\sc Sibelius}, showing the resulting angular correlation functions for different sample sizes of disks and ellipticals in Extended Data Fig.~\ref{fig:assembly}, for the same redshift ranges as in Fig.~\ref{fig:correlation}. When comparing to Fig.~\ref{fig:correlation} it is important to note that because only central galaxies are considered, the correlation of galaxies within the same halo (the so-called one-halo term) is suppressed, resulting in weaker overall correlations.

In every case, as expected, we find a stronger correlation for the central ellipticals than for the central disks. There is some sign of positive assembly bias for our largest samples of galaxies ($N=2000$), but for smaller samples, the assembly bias can appear both positive or negative, and does not appear to be statistically significant.

We have no reason to believe that assembly bias does not also apply in the Local Universe, but it appears that the standard halo bias is the most significant effect in explaining the different distributions of the most massive disks and ellipticals with respect to the supergalactic plane.

\backmatter

\section*{Data availability}
All data used in this work is publicly available. The 2MRS catalogue is provided in \cite{Huchra-2012}. The halo and galaxy catalogues of the {\sc Sibelius Dark} simulation can be accessed via SQL at: \url{https://virgodb.dur.ac.uk}. SQL scripts to obtain the data used in this work are provided with our analysis code.

\section*{Code availability}
The analysis in this paper was performed in python3, and makes extensive use of open-source libraries, including Astropy v.5.1.1 \cite{Astropy}, Matplotlib v.3.7.0 \cite{matplotlib-paper}, NumPy v.1.23.5 \cite{numpy-paper}, SciPy v.1.9.3 \cite{SciPy} and Corrfunc v.2.5.0 \cite{corrfunc-paper}. A documented Juypter notebook containing the code to reproduce all figures in this paper is available at: \url{https://github.com/TillSawala/supergalactic}.

\section*{Acknowledgements}
We sincerely thank P.J.E. Peebles for insightful discussions and guidance and TS thanks S. Räsänen for inspiring discussions. TS and PHJ acknowledge support from the Academy of Finland grant 339127. TS and CSF are supported by the European Research Council (ERC) Advanced Investigator grant DMIDAS (GA 786910) and STFC Consolidated Grant ST/T000244/1. PHJ also acknowledges support from the European Research Council (ERC) Consolidator Grant KETJU (no. 818930). This work was also supported by the Simons Collaboration on ``Learning the Universe''. This work used facilities hosted by the CSC—IT Centre for Science, Finland, and the DiRAC@Durham facility, managed by the Institute for Computational Cosmology on behalf of the STFC DiRAC HPC Facility (www.dirac.ac.uk) and funded by BEIS capital funding via STFC capital grants ST/K00042X/1, ST/P002293/1, ST/R002371/1 and ST/S002502/1, Durham University and STFC operations grant ST/R000832/1. DiRAC is part of the UK National e-Infrastructure. This project used open source software, including Astropy, Matplotlib, Numpy, SciPy and Corrfunc.

\section*{Contributions}
TS and CF developed the original ideas and TS performed the data analysis. JJ and GL performed the reconstruction of the initial conditions for the {\sc Sibelius} simulation, to which TS, CF, JJ, PHJ and GL contributed jointly. TS wrote the first draft, and all co-authors edited and revised the manuscript.

\section*{Competing interests}
The authors declare no competing interests.

\bigskip


\bibliographystyle{naturemag}
\bibliography{main}

\clearpage

\end{document}